# Electron acceleration and X-ray generation from near-critical-density carbon nanotube foams driven by moderately relativistic lasers

Zhuo Pan[1], Jianbo Liu[1], Pengjie Wang[2], Zhusong Mei[1], Zhengxuan Cao[1], Defeng Kong[1], Shirui Xu[1], Zhipeng Liu[1], Yulan Liang[1], Ziyang Peng[1], Tianqi Xu[1], Tan Song[1], Xun Chen[1], Qingfan Wu[1], Yujia Zhang[1], Qihang Han[1], Haoran Chen[1], Jiarui Zhao[1], Ying Gao[1], Shiyou Chen[1], Yanying Zhao[1], Xueqing Yan[1,3,4], Yinren Shou[5*], Wenjun Ma[1,3,4*]

[1]State Key Laboratory of Nuclear Physics and Technology, and Key Laboratory of HEDP of the Ministry of Education, CAPT, Peking University, Beijing 100871, China

[2]Institute of Radiation Physics, Helmholtz-Zentrum Dresden-Rossendorf, Dresden 01328, Germany

[3]Beijing Laser Acceleration Innovation Center, Huairou, Beijing 101400, China

[4]Institute of Guangdong Laser Plasma Technology, Baiyun, Guangzhou 510540, China

[5]Center for Relativistic Laser Science, Institute for Basic Science, Gwangju 61005, Republic of Korea

*Corresponding authors. E-mails: shouyinren@pku.edu.cn; wenjun.ma@pku.edu.cn

**Abstract**

Direct laser acceleration of electrons in near-critical-density (NCD) carbon nanotube foams (CNFs) has its advantages in the high-efficiency generation of relativistic electrons and broadband X-rays. Here, we report the first simultaneous measurement on the spectra of laser-driven electrons and X-rays from CNFs at moderately relativistic intensities of around $5 \times 10^{19}$ W/cm$^2$. The density and thickness of the CNFs were scanned in the experiments, indicating the optimized electrons temperature of 5.5 MeV and X-ray critical energy of 5 keV. Two-dimensional (2D) particle-in-cell (PIC) simulations confirm that the electrons, with a temperature significantly higher than the ponderomotive scale, are directly accelerated by the laser along the NCD plasma channel, while the bright X-rays are emitted by these electrons through betatron radiation or Thomson backscattering inside the channel. The simultaneously generated electrons and X-rays, automatically synchronized with the femtosecond laser driver, are suitable for applications such as bi-modal radiography.

**Introduction**

The advent of chirped pulse amplification[1] has significantly propelled researches in laser-driven acceleration. Today, the high power laser has enabled the generation of GeV-level collimated electrons[2–4] with an energy spread less than 1%[5] through the laser wake field acceleration (LWFA) regime, demonstrating its remarkable success across a range of promising applications. These high-energy electrons, moving in the electromagnetic field within plasmas or undulators, can produce brilliant X-rays with micrometer-scale source sizes and femtosecond-scale durations. Such characteristics are highly pursued in X-ray phase-contrast imaging[6] and ultrafast imaging of high-energy density matter[7]. Therefore, all-optical laser plasma X-ray/gamma-ray sources have become a hot topic. Sources based on LWFA electrons can offer tunable photon energy based on betatron radiation[8–12] or Thomson scattering[13–16]. However, the total charge of electrons generated by the LWFA typically does not exceed hundreds of pC, thus the photon number and energy conversion efficiency are limited[17,18]. SM-LWFA regime could enhance the electron charge by over an order of magnitude[19,20], while it still suffers from the low efficiency and plasma instability. For applications that require high-flux photons, it is necessary to explore other acceleration schemes in which the electron number and the energy conversion efficiency can be significantly promoted.

One of the most promising scheme is the direct laser acceleration in NCD plasma[21,22]. In that regime, a relativistic laser pulse propagates in a NCD plasma[23–25] and forms a micron-scale ion channel. Numerous electrons are confined in the channel and oscillate in the laser and self-generated electromagnetic fields. When the oscillation frequency of electrons in their rest frame matches the laser frequency, the electrons can be resonantly accelerated and continuously gain energy from the laser field. Numbers of theoretical works have predicted that nC, 100s MeV electrons, and ultra-high brightness X-ray/gamma-ray radiations can be generated from the laser NCD plasma interactions by the way of betatron emission[26–28] or Thomson backscattering[29–33]. However, limited experimental results were reported due to the difficulties in the fabrication of NCD targets. One of the earliest experiments using a sub-critical high-pressure gas jet demonstrated DLA electrons with a large divergence angle and Boltzmann-like energy spectrum up to 12 MeV. The energy conversion efficiency from laser to electrons can reach 5%[34]. Kneip et al. measured the electron and X-ray energy spectra from sub-critical gas jets irradiated by 100 J/630 fs laser pulses. They found the optimal electron density for high-energy electron generation is about $10^{19}$ cm$^{-3}$~$0.01n_\text{c}$, where $n_\text{c} = m_e\varepsilon_0\omega_0^2/e^2$ is the critical density. They concluded the measured X-rays up to 50 keV were well described in the synchrotron asymptotic limit of electrons oscillating in a plasma channel[9]. In 2013, Chen et al. utilized 3-TW laser pulses interacting with Ar cluster targets and produced electron beams with charge of up to 200 pC and energy of up to 30 MeV. The measured brightness of the betatron X-rays was $10^{21}$ photons/s/mm$^2$/mrad$^2$/0.1%BW[35]. Rosmej et al. demonstrated that electron beams with charge of up to 100 nC and a laser-to-electron energy conversion efficiency of 30% can be achieved when 100 J/750 fs laser pulses interact with pre-ionized foam targets at the intensity of 2-5× $10^{19}$ W/cm$^2$. High-dose bremsstrahlung radiation was generated when the electrons passed through a high-Z metal target[36,37]. These experiments proved that relativistic laser pulse interaction with NCD plasma in DLA regime could generate high-energy, high-charge electron beams as well as intense X-rays. However, systematic experimental studies with well-controlled target parameters, especially for $n_e \sim n_c$, are rarely reported.

Recently, carbon nanotube foams (CNF) have been successfully introduced as NCD targets in experiments[38]. This foam is synthesized by chemical vapor deposition (CVD) method, formed by intertwined nanometer-thick bundles. Different from conventional foam targets built by micrometer-size blocks, a CNF target can turn into highly homogenized NCD plasma by the rising edge of a laser pulse. Since the thickness and density can be precisely controlled during the fabrication process[39], it is convenient to systematically study and optimize the electron accelerations and X-ray generations. The CNF targets were first used for laser-driven ion acceleration[38,40,41] and got record-breaking enhancement in the ion energy. Thereafter, the generation of brilliant hard x-rays and gamma-rays were demonstrated by utilizing PW lasers at intensity over $10^{21}$ W/cm$^2$ [42]. Its prominent energy conversion efficiency shed a light on the applications requiring a high photon number and large field of view.

In this paper, we present a simultaneous measurement on the spectra of electrons and X-rays from double-layer carbon nanotube targets irradiated by femtosecond pulses at moderate relativistic intensity of $5 \times 10^{19}$ W/cm$^2$. By varying the densities and thicknesses of CNF layer, the dependence of electron temperature and X-ray photon energy was studied. 2D PIC simulations with experimental parameters were performed to illustrate the electron acceleration and X-ray emitting process. The simulated electron spectra, and the derived betatron and Compton scattering spectra, fit well to the experimental results.

**Experiment Setup and Results**

The experiments were performed at Peking University using the Ti: sapphire laser system CLAPA. In the campaign, 30 fs/1.0 J pulses with linear polarization were delivered on targets. By using an f/3.5 off-axis parabola mirror (OAP), the spot size is measured as 5.0 μm full width at half maximum (FWHM) with 30% energy concentration[43]. The laser peak intensity was $4.4 \times 10^{19}$ W/cm², corresponding to a normalized vector potential amplitude of $a_0 = eE/m_e c\omega \approx 4.5$. The double-layer targets are composed of carbon nanotube foams (CNFs) gaplessly attached in front of a thin plastic foil. The CNF, characterized by 10-nm diameter filaments that are intricately interwoven, is synthesized via the chemical vapor deposition method.

The density of the CNFs varying from 1.1 mg/cm³ to 4.0 mg/cm³, corresponding to 0.2 to 0.7 $n_c$ for the wavelength of 800 nm. We choose the thickness varying from 40-80 μm for $0.2n_c$ and 20-60 μm for $0.7n_c$ target. Thanks to the cross-polarized wave front-end and a single plasma mirror system, the laser contrast ratio is higher than $10^{10}$ until 40 picoseconds before the main pulse, preventing the premature damage of targets. The schematic setup is illustrated in Fig. 1. The electrons and X-rays driven by the relativistic laser passed through a Teflon collimator of 2-mm thickness and 3-mm diameter positioned 30 cm away from the targets on the laser axis. The spectrum of electrons was measured by a dipole magnet of 0.29 T in conjunction with a scintillator (Biomax, Kodak) imaged by an Andor CCD. The X-rays were measured by Fuji-Film MS imaging plates (IPs) 67 cm away from the targets outside the chamber after a 300-μm-thick Be window. A cake filter made of three different metal foils (20 μm Al, 20 μm Cu, 40 μm Cu) was put between the IP and the Be window constituting a four-petal filter to measure the x-ray spectrum. The drift distance of 24 cm from the magnet to the IPs guaranteed that electrons below 80 MeV would not hit on the IPs.

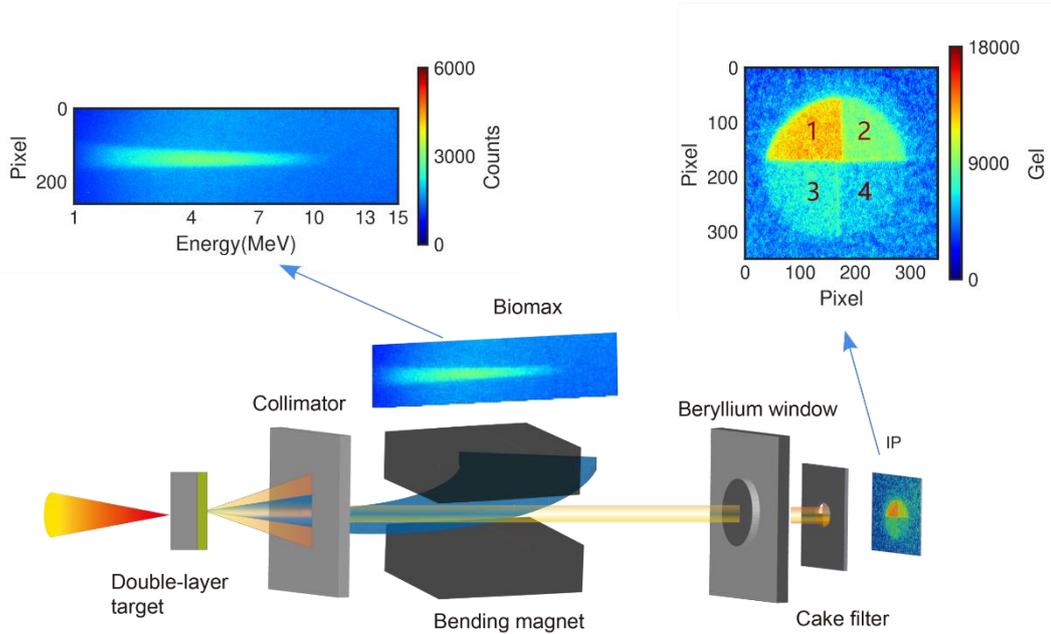

Fig. 1. Schematic of the experiment setup. Left inset shows the electron spectrum recorded by the Biomax screen, which is captured by an CCD. Right inset shows the X-ray profile imaged by an imaging plate after the X-ray filter.

To derive the electron spectrum from the spectrometer, we first acquired the distribution profile of the magnetic field by CST simulations[44]. Subsequently, a correlation map could be established between the energy of the incident particles and their corresponding position on the recording media. With this correlation map in place, it became feasible to infer the electron spectrum from the observed data. Figure 2(a) shows the derived electron spectrum for a 60 μm/$0.2n_c$ CNF attached with a 0.5-μm-thick CH target. The spectrum has a Boltzmann-like distribution with a fitted temperature of 3.6 MeV. We performed a series of experiments to study the dependence of electron and X-ray spectra on the target parameters by varying the CNF's thickness (L) and electron density ($n_e$). All the electron spectra in the measurement range have the Boltzmann-like distribution, but the numbers and temperatures strongly rely on the target parameters. The $T_e$ of $0.2n_c$ targets are significantly higher than that of $0.7n_c$ targets by a factor of 3-5, maximized at L=80 μm with $T_e = 5.5$ MeV, and the electron number of $0.2n_c$ targets are also higher. For targets with $n_e = 0.2\ n_c$, the temperature and number of electrons increases with the thickness L. While in the targets of higher density of $0.7n_c$, the dependence of $T_e$ on the thickness of the targets is weak, and the variance of the fitted temperature is no more than 10%.

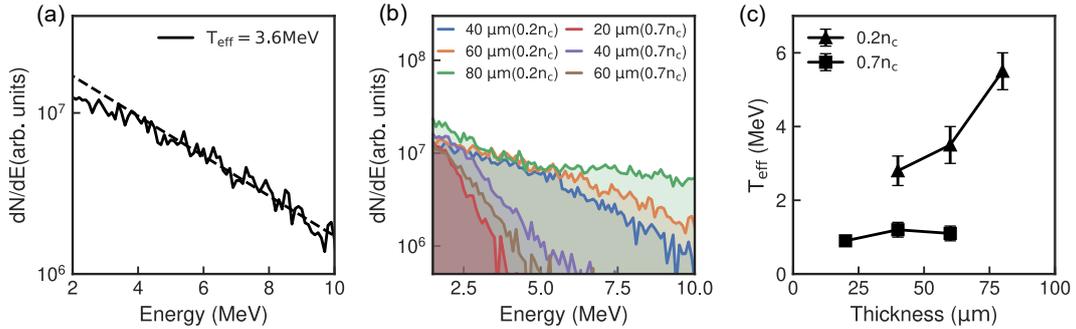

Fig. 2. Experimental results of electron spectrum. (a) A typical experiment result obtained from a 60 μm CNF (0.2 $n_c$) + 0.5 μm CH target. (b) Measured energy spectra of the electrons with different target parameters. (c) Electron temperatures with varied densities and thicknesses.

The four-petal metal filters made of Be, Cu, and Al foils were used to measure the X-ray spectrum along the laser direction. We calculate the transmissions of different filters listed in Fig. 3(a) with the software XOP[45]. The filters attenuate the X-ray photon flux. By using multiple filters, the X-ray spectrum can be resolved through the combination of the transmission curve and the integral response. We define the cut-off energy of the filter by the energy where the transmission is 1%. The corresponding cut-off energy for filter No.1-4 is about 2.5, 3.6, 4.7 and 6.5 keV, respectively. One can see from Fig.1 that the X-ray signals drop dramatically for the 4th filter, which indicates that the X-ray photons are mainly below 6.5 keV. Figure 3(b) shows the photostimulable luminescence (PSL) data of the X-rays measured by the IPs behind the cake filter. In the case of betatron radiation, the synchrotron-like radiation spectrum can be described as[17] $S(x) = x \int_x^\infty K_{5/3}(\xi) d\xi$, Where $K_{5/3}$ is a modified Bessel function of the second kind, $x = \omega/\omega_c$, and $\omega_c$ is the critical energy. The X-ray spectrum can be fitted with this shape to match the experimental results by the least-squares fit and the critical energy can be derived from the fitting function. The fitted critical energy is 3.5 keV, which is displayed in Fig. 3(b). By integrating the spectra ranging from 2-10 keV, an X-ray yield of $1.4 \times 10^9$ photons/Sr can be obtained. Figures 3(c) and (d) show the X-ray data from the 0.2 and $0.7n_c$ targets with different thicknesses. The photon numbers from the $0.7n_c$ targets are higher than that from the $0.2n_c$ targets. However, the critical photon energy $E_c$ of the $0.2\ n_c$ targets is higher instead. For targets with $n_e = 0.7n_c$, the photon number and $E_c$ is almost independent on the thickness of the targets. While for $0.2n_c$ targets, the minimum $E_c$ is achieved for L = 60 μm, and the highest $E_c$ of 5 keV is obtained for L = 80 μm.

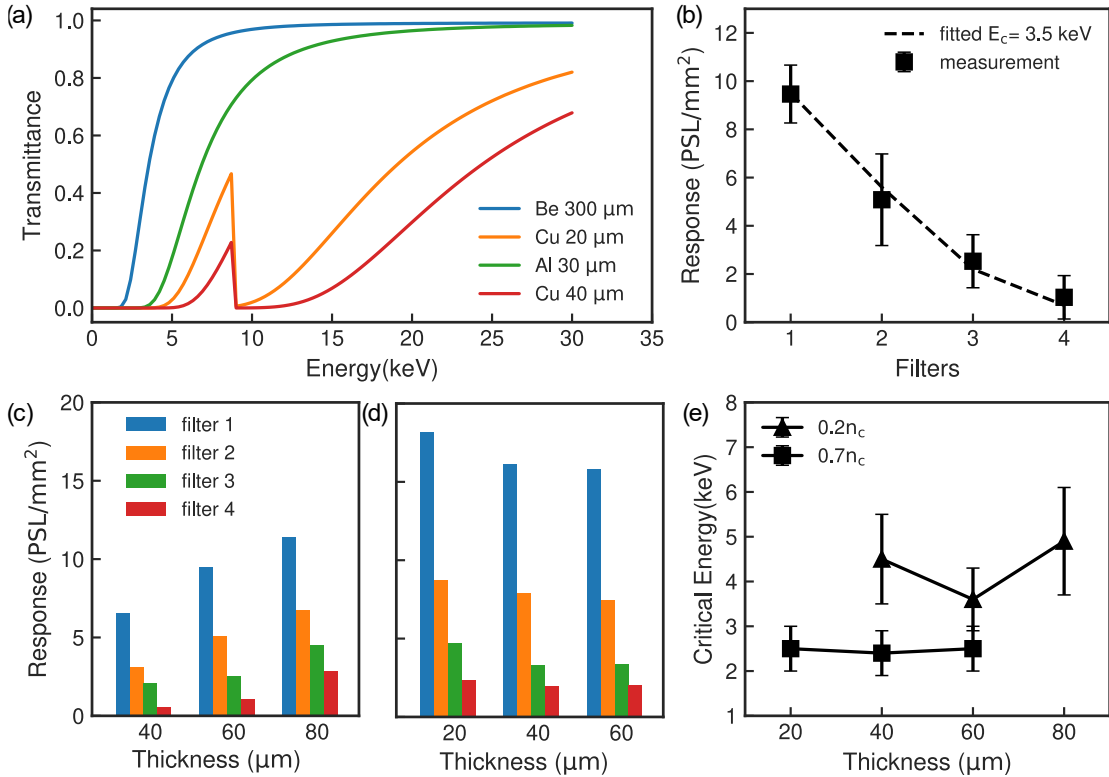

Fig. 3. Experimental results of X-ray spectrum (a)Transmissions of different metal filters. (b) Experiment result obtained from a 60 μm CNF ($0.2n_c$) + 0.5 μm CH target and synchrotron model fitted critical energy. (c) Responses from IP for $0.2n_c$ targets

and (d) for $0.7n_c$ targets with different target thicknesses. (e) The fitted critical energies with varied densities and thicknesses.

**Discussions and Simulations**

Upon examination of the experimental data, the following interpretations can be drawn. First, the experimental results unambiguously indicate that $n_e = 0.2n_c$ targets outperform those with an electron density of $0.7n_c$ in terms of the generation of high-energy electrons at $a_0 \sim 5$. Previous theories predicted that [32, 33] electrons in the NCD channel gain the highest energy when $G = a_0\sqrt{n_e/n_c} \sim 1$. In our case, the G for $0.2\ n_c$ and $0.7n_c$ target is 2.2 and 4.2, respectively. So, the result of higher $T_e$ for $0.2n_c$ is in consistent with the theories. Second, for the $0.7n_c$ targets, the $T_e$, $E_c$ and X-ray yield are similar for L = 20, 40 and 60 μm, which indicates their electron acceleration and X-ray emission process are similar. Previous studies reveal that the depletion length (This describes the distance over which the total energy of the laser would be absorbed.) of laser in plasma $L_{etch} \approx \frac{1}{4}\frac{a_0}{n_e/n_c}c\tau$ [31,47]. Calculated from this formula, $L_{etch} = $ 15 μm for $n_e = 0.7\ n_c$. It means the laser pulses would deplete their energy after propagating over 15 μm in $0.7n_c$ CNF. This explains why the electron and X-ray spectra are similar for CNF with L=20-60 μm. Third, the different correlation between the spectra of the electrons and the X-rays suggests the dominant X-ray emission processes in the two kinds of targets are different. For the $0.7n_c$ targets, both electron temperature $T_e$ and the photon critical energy $E_c$ show weak relevance of the target thickness. For the $0.2n_c$ targets, however, the $T_e$ rises with increasing L. while the $E_c$ is minimal for L=60 μm.

To confirm our interpretations and reveal the underlying physics, we first performed a 2D PIC simulations with the SMILEI code[48] for the 60 μm/$0.2n_c$ target. The simulation utilized a fixed window of 100 μm in the laser propagation direction and 24 μm in the transverse direction. The longitudinal and transverse resolution is $\lambda/64$ and $\lambda/16$, respectively. The driving laser's normalized amplitude $a_0$, duration τ, and laser wavelength $\lambda_L$, waist size $w_0$, was set as 4, 30 fs, 0.8 μm, 4.2 μm, respectively, to match the experimental parameters. A $200\ n_c/0.5$ μm plasma is positioned at 70 μm to represent the CH foil. A uniform carbon plasma with electron densities of $0.2n_c$ is positioned from 10 μm to 70 μm to represent the CNF targets. The position of the laser focus is set on the interface of two layers. The trajectories of test electrons are also tracked.

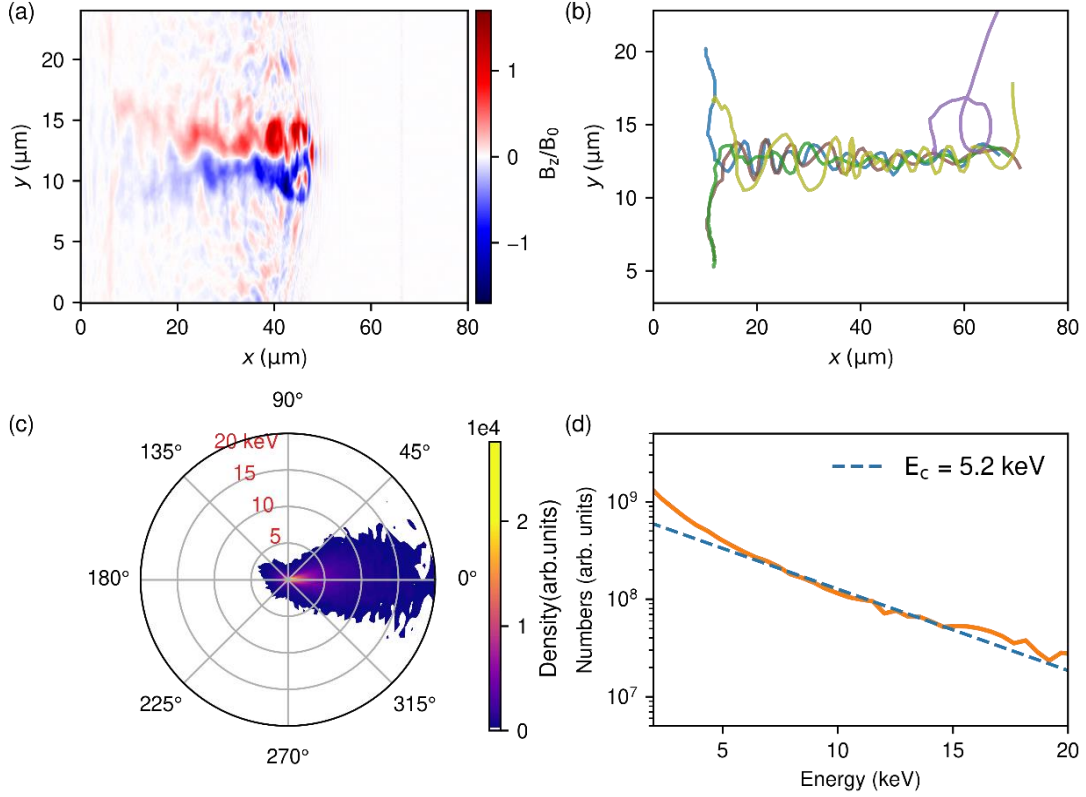

Fig. 4 (a) $B_z$ field distribution at t = 50T which shows evident static magnetic field structures. (b) Trajectories of several selected electrons from the plasma background which are accelerated by the laser field and oscillate in the plasma channel. (c) Photon density distribution at t = 90T. (d) Photon spectrum from the simulation at t = 90T and the fitting critical energy which agrees well with the experiments.

As the laser travelling in the near critical density plasma, the electrons are pushed away by the laser pondermotive force, leading to the formation of a plasma channel. The self-generated static electric and magnetic field confine the energetic electrons which will experience transverse betatron oscillations in the channel. In Fig. 4(a) we can observe the toroidal static magnetic field, which is the distinguishing feature of the direct laser acceleration, is formed at 50T (where T is the laser period). The magnetic field can reach to $B_z = 1.7 B_0$ which has the ability to confine the motion of the electrons. Here $B_0 = 1.35 \times 10^4$ T. We select several representative electrons depicting their evolution of trajectories from 10T to 70T, as shown in Fig. 4(b). The electrons are driven from the plasma background and then injected into the channel, keeping their motion confined in the channel for over 40 μm. The electrons gradually gain energy from the laser field and finally become the superpondermotive electrons.

When the electrons travel in the plasma channel and are oscillated by the laser pulse and self-generated static field, enormous photons would be emitted from this process by betatron radiation regime. The built-in method of the SMILEI code could help us to simulate the generation of keV-level photons alongside the electron acceleration. The photon spectrum and the angular distribution of the photons are diagnosed. Figure 4(c) indicates that the photons have the property of a forward direction and the full width at half maximum (FWHM) angle is about 28°. The critical photon energy obtained by fitting with the synchrotron model is displayed in Fig. 4(d), which fits the experimental results well.

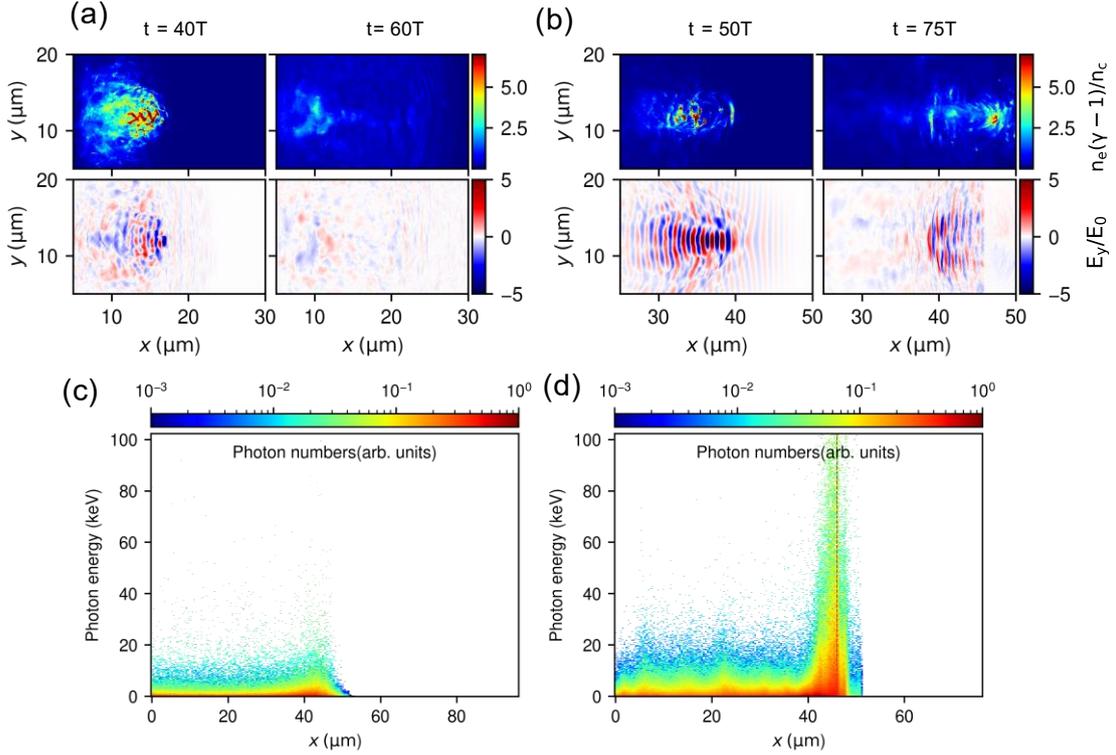

Fig 5. (a) The energy density of electrons and electric field for the case of 60μm/$0.7n_c$ at different times. (b) The same values for the case of 40 μm/$0.2n_c$. (c) and (d) show the photon spectrum along the laser axis at the time of 70T, when the laser pulse arrives at the second layer. The dashed red line marks the interface of the two layers.

To explore the details of various thickness and density settings, we also perform a series of simulations. We set the thickness of 20/40/60 μm with density of $0.7n_c$, and 40/60/80 μm with $0.2\,n_c$ in double-layer targets while keeping other parameters unchanged. For the higher density plasma, snapshots of the electric field $E_y$ and the energy density $n_e(\gamma - 1)/n_c$ at separate times are presented in Fig. 5(a). At 40T, the laser pulse has a steeper rising edge and smaller focus size due to the self-focusing. The accelerated electrons follow the laser. while at the 60T, the laser pulse has almost depleted in the plasmas and the $E_y$ field almost vanishes. We conclude that the laser propagation length in $0.7n_c$ plasma is less than 20 μm, which agrees well with the prediction on the laser penetration length, suggesting that there is only betatron emission from DLA electrons. While for the $0.2n_c$ case, at 50T the self-steepening and self-focusing of the laser pulse still happen in the near critical density plasma. The normalized electric field can attain a value of 7.74, which corresponding to the intensity of $9.3 \times 10^{19}$ W/cm². When compared with the initial intensity of $2.2 \times 10^{19}$ W/cm² as the laser pulse enters the plasma from left boundary, this represents an enhancement in the intensity by a factor of 4.2 times. The pondermotive force builds a plasma channel and the laser is followed by a series of high energy bunches. At 75T the laser pulse is reflected by the second layer and interacts with the electrons behind, generating the photons by the collision process. Different from the betatron emission, the Compton scattering regime boost the photon energy by $4\gamma^2$, and is synchronized in temporal and spatial dimension. Both the enhanced laser intensity and higher electron energy could increase the energy of the photon from this regime. As the thickness increases, the laser absorption becomes more efficient. When the thickness is increased to 80 μm, the laser pulse is nearly totally absorbed by the plasma, changing the regime of photon generation to betatron emission which is like the 0.7nc case.

The photon spectrum along the laser axis is presented in Fig. 5(c) and (d), for the target of 60 μm/$0.7n_c$ CNF, the DLA electrons are confined by the self-generated magnetic field, leading the photons no more than 20 keV which are based on synchrotron emission regime, while for the target of 40 μm/$0.2n_c$ CNF, at the interface position of the two layers, the laser reflects and collides with the electrons, increasing the photon spectrum up to 100 keV.

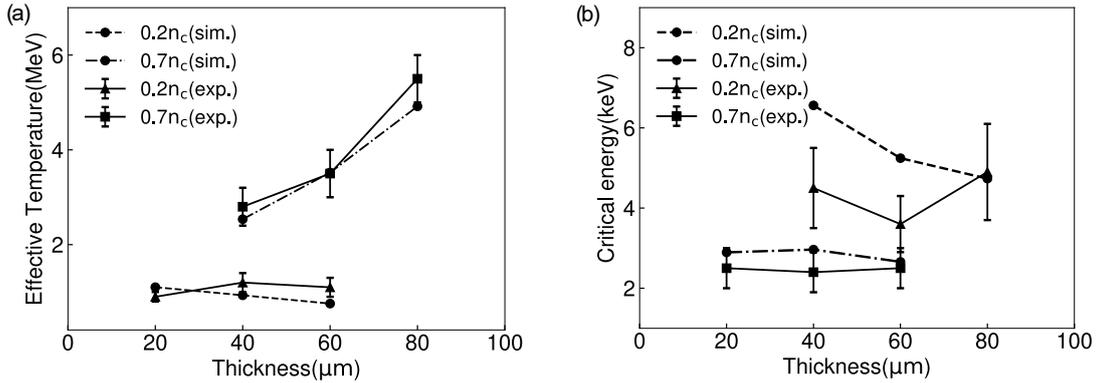

Fig. 6 Comparison of simulated and the experimental results. (a) The effective temperature of electrons and (b) the critical energy of photons. The dashed lines are obtained from PIC simulations and the solid lines are obtained from the experimental fitting results.

From the series of simulations, the effective temperature of electrons and the critical energy of photons are compared in Fig. (6). In the high-density plasma, both the effective temperature and the critical energy are essentially invariant to the thickness of the target which agrees with the observed one from the experiments. While in the low-density plasma, the effective temperature of the electrons shows the increasing trends with thickness which fits the experimental results very well, and the critical energy of the photons show a clear increase for the short thickness due to the transition from the betatron emission regime to the Compton scattering regime.

**Conclusion**

In summary, we simultaneously measured the electron and the X-ray energy spectra from laser-irradiated $0.2n_c$ and $0.7n_c$ near-critical-density double-layer target at intensity of around $5 \times 10^{19}$ W/cm$^2$. Electrons with temperatures of 1-5.5 MeV, and X-rays with critical energies of 2.5-5 keV were obtained depending on the target parameters. The greatest brilliance is achieved with a target of 20 μm thickness and 0.7 $n_c$ density. The 2D PIC simulations based on the experimental parameters confirm that the super-ponderomotive electrons in our targets are accelerated through direct laser acceleration in a NCD plasma channel. The betatron motion of the electrons universally leads to the X-rays emission. By reducing the density of CNF, the dominant radiation process switches from betatron emission to Compton scattering. Our results provide a fresh and complete dataset for the study of the electron acceleration and X-ray generation in NCD plasma at relativistic intensity, which has the potential for applications in electron imaging and X-ray imaging such as bi-modal radiography.

**Acknowledgements**

This work was supported by the following projects: NSFC Innovation Group Project (grant number 11921006) and National Grand Instrument Project (grant number 2019YFF01014402). W. Ma acknowledges support from the National Science Fund for Distinguished Young Scholars (12225501). The PIC simulations were carried out on High-performance Computing Platform of Peking University.

**Author Declarations**

**Conflict of Interest**

The authors have no conflicts to disclose.

**Data Availability**

The data that support the findings of this study are available from the corresponding author upon reasonable request.